\documentclass[arXiv]{actapoly}

\usepackage{color}
\usepackage{xspace}
\usepackage{graphicx}
\usepackage{natbib}
\usepackage{amsmath}
\usepackage{microtype}

\begin{document}

\title{The Goodness of Simultaneous Fits in ISIS}

\correspondingauthor[M. K\"uhnel]{Matthias K\"uhnel}{a}{matthias.kuehnel@sternwarte.uni-erlangen.de}
\author[S. Falkner]{Sebastian Falkner}{a}
\author[C. Grossberger]{Christoph Grossberger}{a}
\author[R. Ballhausen]{Ralf Ballhausen}{a}
\author[T. Dauser]{Thomas Dauser}{a}
\author[F.-W. Schwarm]{Fritz-Walter Schwarm}{a}
\author[I. Kreykenbohm]{Ingo Kreykenbohm}{a}
\author[M.~A. Nowak]{Michael~A. Nowak}{b}
\author[K. Pottschmidt]{Katja Pottschmidt}{c,d}
\author[C. Ferrigno]{Carlo Ferrigno}{e}
\author[R. E. Rothschild]{Richard E. Rothschild}{f}
\author[S. Mart\'inez-N\'u\~nez]{Silvia  Mart\'inez-N\'u\~nez}{g}
\author[J. M. Torrej\'on]{Jos\'e Miguel Torrej\'on}{g}
\author[F. F\"urst]{Felix F\"urst}{h}
\author[D. Klochkov]{Dmitry Klochkov}{i}
\author[R. Staubert]{R\"udiger Staubert}{i}
\author[P. Kretschmar]{Peter Kretschmar}{j}
\author[J. Wilms]{J\"orn Wilms}{a}

\institution{a}{Remeis-Observatory \& ECAP, Universit\"at Erlangen-N\"urnberg, Sternwartstr. 7, 96049 Bamberg, Germany}
\institution{b}{MIT Kavli Institute for Astrophysics, Cambridge, MA 02139, USA}
\institution{c}{CRESST, Center for Space Science and Technology, UMBC, Baltimore, MD 21250, USA}
\institution{d}{NASA Goddard Space Flight Center, Greenbelt, MD 20771, USA}
\institution{e}{ISDC Data Center for Astrophysics, Chemin d'\'Ecogia 16, 1290 Versoix, Switzerland}
\institution{f}{Center for Astronomy and Space Sciences, University of California, San Diego, La Jolla, CA 92093, USA}
\institution{g}{X-ray Astronomy Group, University of Alicante, Spain}
\institution{h}{Cahill Center for Astronomy and Astrophysics, CALTECH, Pasadena, CA 91125, USA}
\institution{i}{Institut f\"ur Astronomie und Astrophysik, Universit\"at T\"ubingen, Sand 1, 72076 T\"ubingen, Germany}
\institution{j}{European Space Astronomy Centre (ESA/ESAC), Science Operations Department, Villanueva de la Ca\~nada (Madrid), Spain}

\begin{abstract}
In a previous work, we introduced a tool for analyzing multiple datasets
simultaneously, which has been implemented into ISIS. This tool was used to
fit many spectra of X-ray binaries. However, the large number of degrees of
freedom and individual datasets raise an issue about a good measure for a
simultaneous fit quality.

We present three ways to check the goodness of these fits: we investigate the
goodness of each fit in all datasets, we define a combined goodness exploiting
the logical structure of a simultaneous fit, and we stack the fit residuals of
all datasets to detect weak features. These tools are applied to all
\textsl{RXTE}-spectra from GRO~1008$-$57, revealing calibration features that
are not detected significantly in any single spectrum. Stacking the
residuals from the best-fit model for the Vela~X-1 and XTE~J1859+083 data
evidences fluorescent emission lines that would have gone undetected
otherwise.
\end{abstract}


\maketitle

\section{Introduction}

\begin{figure}
  \centering
  \includegraphics[width=.9\linewidth]{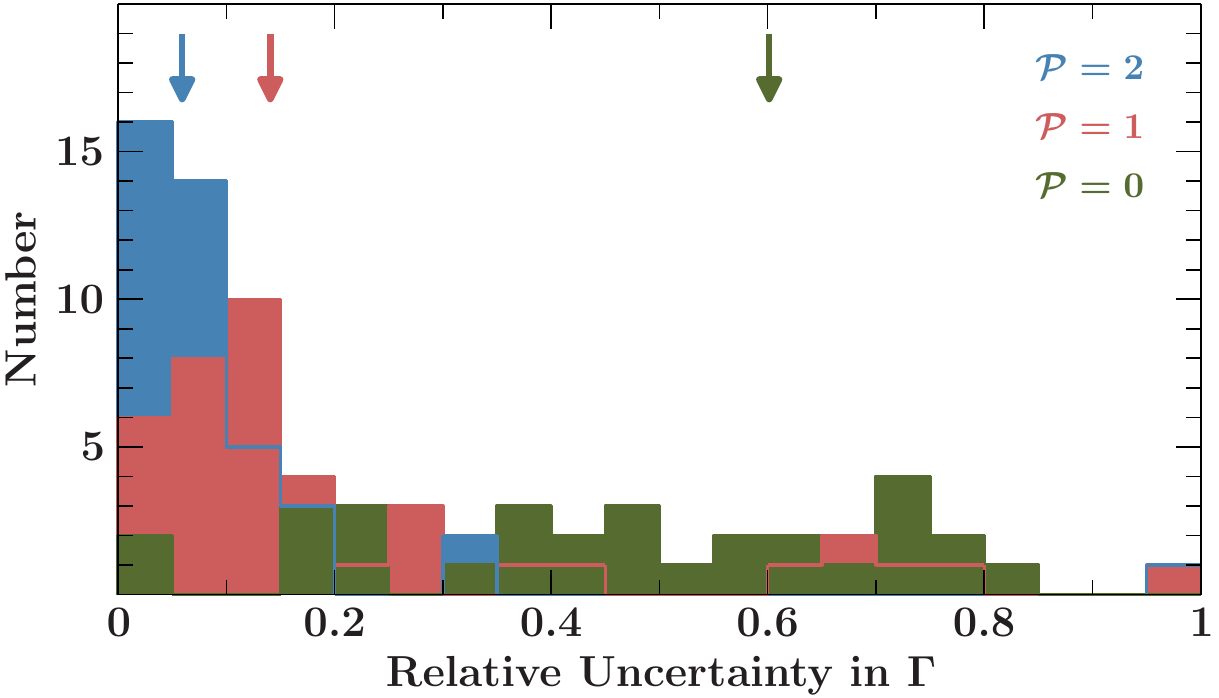}
  \caption{Distribution of the relative uncertainties (90\% confidence level)
  of the power-law photon indices, $\Gamma$, in our study of all
  \textsl{RXTE}-observations of GRO~J1008$-$57 \citep{kuehnel2013a}. Fitting
  all 43 observations separately results in the green histogram. As soon as we
  perform a simultaneous fit with $\mathcal{P}=1$ global continuum parameter
  the uncertainties decrease significantly as shown by the red histogram.
  Finally, using $\mathcal{P}=2$ global parameters (blue histogram) results in
  a median of ${\sim}6\%$ in the uncertainties (compare the arrows on top).}
  \label{fig:1008gamma}
\end{figure}

Nowadays, the still increasing computation speed and available memory allows
us to analyze large datasets at the same time. Using X-ray spectra of
accreting neutron stars as an example, we have shown in a previous paper
\citep[][hereafter paper I]{kuehnel2015a} that loading and fitting the spectra
simultaneously has several advantages compared to the ``classical'' way of
X-ray data analysis, which is treating every observation individually. In
particular, instead of fixing parameters to a mean value one can determine
them by a joint fit to all datasets under consideration. Due to the
reduced number of degrees of freedom the remaining parameters can be better
constrained (see Fig.~\ref{fig:1008gamma} as an example). Furthermore,
parameters no longer need to be independent, but can be combined into
functions. For instance, the slope of the spectra might be described as a
function of flux with the coefficients of this function as fit-parameters.

The disadvantages of fitting many datasets simultaneously are, however, an
increased runtime and a complex handling because of the large number of
parameters. In paper I, we have introduced functions to facilitate this
handling, which have been implemented into the \texttt{Interactive Spectral
Interpretation System} (ISIS) \citep{houck2000a}. While these functions are
already available as part of the
\texttt{ISISscripts}\footnote{http://www.sternwarte.uni-erlangen.de/isis} they
are continuously updated and new features are implemented. One important
question, which we raised in paper I, is about the goodness of a
simultaneous fit as it is, e.g., calculated after the commonly used
$\chi^2$-statistics, particularly the case where some datasets are
not described well by the chosen model. Due to the potential large total
number of datasets, the information about failed fits can be
hidden in the applied fit-statistics. After we have given a reminder
about the terminology of simultaneous fits in Section~\ref{sec:reminder},
we describe the problem of detecting failed fits in more detail in
Section~\ref{sec:solution} and provide possible solutions. We will
conclude this paper by applying these solutions to examples in
Section~\ref{sec:examples}.

\subsection{Simultaneous Fits}
\label{sec:reminder}

\begin{figure}
  \centering
  \includegraphics[width=.9\linewidth]{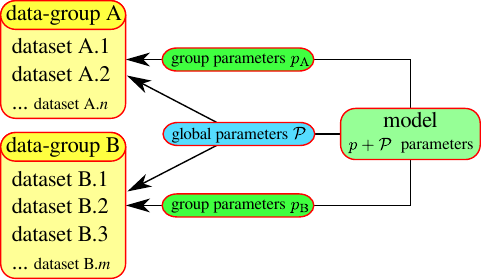}
  \caption{Terminology of simultaneous fits in ISIS, according to paper I. A
  data-group consists of simultaneously taken datasets, here data-group A has $n$
  and B has $m$ datasets. A model with $p+\mathcal{P}$ parameters is fitted
  such that each data-group has its own set of $p_i$ parameters, called group
  parameters. The common $\mathcal{P}$ parameters between the groups are the
  so-called global parameters.}
  \label{fig:terminology}
\end{figure}

As we have described in paper I, a \textit{data-group} contains all datasets
which have been taken simultaneously in time or, in general, represent the
same state of the observed object. In the example illustrated in
Fig.~\ref{fig:terminology} two data-groups, A and B, have been added to the
simultaneous fit, containing $n$ and $m$ datasets, respectively. Thus, a
dataset is labeled by the data-group it belongs to, e.g., B3 is the third
dataset in the second data-group.  After a model with $p$ parameters has been
defined each of these data-groups is fitted by an individual set of parameters,
called \textit{group parameters}. Consequently, all datasets belonging to a
specific data-group are described by the same parameter values. A specific group
parameter can now be marked as a so-called \textit{global parameter}. The
value of the corresponding group parameters will now be tied to this global
parameter, i.e, this parameter has a common value among all data-groups.
Instead of tying group parameters together to a global value, a parameter
function may be defined, to which the group parameters are set instead. This
function takes, e.g., other parameters as input to calculate the value for
each group parameter. In this case, correlations between model parameters,
e.g, as predicted by theory can be implemented and fitted directly.

\section{Goodness of a simultaneous fit}
\label{sec:solution}

As an indicator for the goodness of a fit the analysis software used,
e.g ISIS or XSPEC \citep{arnaud1996a}, usually displays the fit-statistics
after the model has been fitted to the data. Here, we chose the
$\chi^2$-statistics since the developed functions for a simultaneous fit have
been first applied to accreting neutron stars. The high count rates satisfies
the Gaussian approxmation of the uncertainties, which are actually Poisson
distributed. In principle, however, the discussed issues and their solutions
can be generalized for any kind of fit-statistics.

For each datapoint $k$ the difference between the $\mathrm{data}$ and the
$\mathrm{model}$ is calculated and normalized by the measurement uncertainty,
$\mathrm{error}$, of the data. The sum over all $n$ datapoints is called the
$\chi$-square,
\begin{equation}\label{eq:chisqr}
  \chi^2 = \sum_{k=1}^n \frac{(\mathrm{data}_k -
    \mathrm{model}_k)^2}{\mathrm{error}_k^2}
\end{equation}
and is displayed after a fit. Additionally, the sum is normalized to the
total number of degrees of freedom, $n-p$ with the number of free
fit-parameters $p$, since the $\chi^2$ increases with $n$. This normalized sum,
called the reduced $\chi$-square,
\begin{equation}\label{eq:redchisqr}
  \chi^2_\mathrm{red} = \frac{\chi^2}{n-p}
\end{equation}
is also displayed. For Gaussian distributed data the expected value
is $\chi^2_\mathrm{red}=1$ for a perfect fit of the chosen model. However,
once the probability distribution is changed, e.g., when a
spectrum has been rebinned, the expected value changes as well. Consequently,
a reliable measure for the goodness of the fit has to be defined with some
forethought.

The $\chi^2_\mathrm{red}$ threshold, for which a simultaneous fit is
acceptable, strongly depends on the considered case. In particular, a few
data-groups might not be described well by the chosen model, which would
result in an unacceptable $\chi^2_\mathrm{red}$ when fitted individually.
However, in case of a simultaneous fit, this information might be hidden in
the classical definition of the $\chi^2_\mathrm{red}$ (Eq.~\ref{eq:redchisqr}). 
Let us consider $N$ data-groups and a model with $p$
group parameters and $\mathcal{P}$ global parameters. Then, the total
$\chi^2_\mathrm{red}$ is
\begin{equation}\label{eq:redchisqr:simfit}
  \chi^2_\mathrm{red} = \frac{\sum_{i=1}^N \chi^2_i}{\sum_{i=1}^N (n_i -
  p_i) - \mathcal{P}}
\end{equation}
with the number of degrees of freedom, $n_i - p_i$, and the $\chi^2_i$ for
each data-group $i$ after Eq.~\ref{eq:chisqr}. Now, we assume a failed fit
with $\chi^2_i \sim 2$ for a particular $i$ to be present, while for the
remaining data-groups $\chi^2_i \sim 1$. For $N \gtrsim 10$ the
$\chi^2_\mathrm{red}$ after Eq.~\ref{eq:redchisqr:simfit} is still near unity
and, thus, suggests a successful simultaneous fit. In the
following, we present three possibilities to investigate the
goodness of a simultaneous fit more carefully.

\subsection{Histogram of the goodness}
\label{sec:histogram}

A trivial but effective solution is to check the goodness of the fit for
each data-group individually. Here, in the chosen case of the
$\chi^2$-statistics, the $\chi^2_\mathrm{red,i}$ is calculated for each
data-group, $i$, after
\begin{equation}\label{eq:groupchi}
  \chi^2_\mathrm{red,i} = \frac{\chi^2_i}{n_i - p_i}
\end{equation}
where $n_i$ are the number of datapoints in the data-group, and $p_i$ is the
number of free group parameters. Due to that the global parameters are
not taken into account here, the $\chi^2_\mathrm{red,i}$ is, however,
different to that performed by a single fit of the data-group.

In the case of a large number of data-groups, it is more convenient to sort the
$\chi^2_\mathrm{red,i}$ into a histogram to help investigating the goodness
of the fit to all data-groups. We have added such a histogram to the
simultaneous fit functions as part of the \texttt{ISISscripts}. After a fit
has been performed using the fit-functions \texttt{fit\_groups} or
\texttt{fit\_global} (see paper I) this histogram is added to the default
output of the fit-statistics. In this way, failed fits of specific data-groups
can be identified by the user at first glance.

\subsection{A combined goodness of the fit}
\label{sec:newstatistic}

Instead of a few failed fits to certain data-groups, one might ask if the
chosen model fails in the global context of a simultaneous fit. To
answer this question, a special goodness of the simultaneous fit is needed to
take its logical structure into account. As explained in 
Section~\ref{sec:reminder}, a data-group represents a certain state of the
observed object, e.g., the datasets where taken at the same time. Thus, the
data-groups are statistically independent of each other. Calculating the
goodness of the fit in a traditional way, which is the $\chi^2_\mathrm{red}$
after Eq.~\ref{eq:redchisqr:simfit} in our case, does not, however, take
this aspect into account. As a solution we propose to define a
\textit{combined goodness of the fit} calculating the weighted mean of the
individual goodness of each data-group. In the case of $\chi^2$-statistics, a
combined reduced $\chi^2$ is calculated by
\begin{equation}\label{eq:combredchisqr}
  \chi^2_\mathrm{red,comb.} = \frac{1}{N} \sum_{i=1}^N \frac{\chi_i^2}{n_i
  - p_i - \mu_i \mathcal{P}} 
\end{equation}
with $\chi_i^2$ computed after Eq.~\ref{eq:chisqr} for each data-group, $i$,
and a weighting factor, $\mu_i$, for the number of global parameters,
$\mathcal{P}$:
\begin{equation}\label{eq:mu}
  \mu_i \approx (n_i - p_i) \times \sum_{j = 1}^N \frac{1}{n_j - p_j}
\end{equation}
Thus, $\mu_i$ normalizes the effect of data-group $i$ on the determination of
the global parameters, $\mathcal{P}$, by its number of degrees of freedom
relative to the total number of degrees of freedom of the simultaneous fit.
Equation~\ref{eq:mu} is, however, an approximation only. A data-group might not
be sensitive to a certain global parameter, e.g, if the spectra in this
data-group do not cover the energy range necessary to determine the parameter.

A failed fit to a specific data-group, for example with a high individual
$\chi^2_\mathrm{red,i}$, has a higher impact on the $\chi^2_\mathrm{red,comb.}$
(Eq.~\ref{eq:combredchisqr}) than on the traditional
$\chi^2_\mathrm{red}$ (Eq.~\ref{eq:redchisqr}). In general we expect
$\chi^2_\mathrm{red,comb.} \geq \chi^2_\mathrm{red}$, even if all data-groups
are fitted well. In the case of a good simultaneous fit (better than a certain
threshold), a weak feature in the data might still be unnoticed, if it is not
detected in any individual data-group. Such a feature can be investigated by
stacking the residuals, as outlined in the following section.

We note, however, that Eq.~\ref{eq:combredchisqr} is the result of an
empirical study. A more sophisticated goodness of a simultaneous fit
should be based on a different type of fit-statistics suitable for a
simultaneous analysis of many datasets, such as a Bayesian approach or a
joint likelihood formalism similar to \citet{anderson2015a}.

\subsection{Stacked residuals}
\label{sec:summedresiduals}


Once datasets can be technically stacked to achieve a higher
signal-to-noise ratio, e.g., when spectra have the same energy grid and
channel binning, further weak features might get visible. This is a common
technique in astrophysics \citep[see, e.g.,][]{ricci2001a,bulbul2014a}.
However, when stacked datasets are analyzed, differences in the individual
datasets, like source intrinsic variability, can no longer be revealed.

In case of a simultaneous fit, the residuals of all data-groups can be
stacked instead. The stacking dramatically increases the total exposure in
each channel bin. Thus, the stacked residuals of all data-groups,
$R(k)$, as a function of the energy bin, $k$, can be investigated to further
verify the goodness of the simultaneous fit
\begin{equation}\label{eq:sumresiduals}
  R(k) = \sum_{i=1}^N \mathrm{data}_{i,k} - \mathrm{model}_{i,k}
\end{equation}
This task can be achieved using, e.g, the \texttt{plot\_data}
function\footnote{\url{http://space.mit.edu/home/mnowak/isis\_vs\_xspec/plots.html},
which is available through the \texttt{ISISscripts} as well.} written by
M.~A.~Nowak.

We can show that the combined reduced $\chi^2$ is effectively equal
to the goodness of a fit of the stacked data in the first place. Assuming the
same number of degrees of freedom, $n-p$, for each data-group,
Eq.~\ref{eq:combredchisqr} gives
\begin{equation}\label{eq:chisqrequal}
  \chi^2_\mathrm{red,comb.} = \frac{1}{f} \sum_{i=1}^N
  \sum_{k=1}^{n} \frac{(\mathrm{data}_{i,k} -
    \mathrm{model}_{i,k})^2}{\mathrm{error}_{i,k}^2}
\end{equation}
with $f = N (n-p-\mu\mathcal{P})$ and having used Eq.~\ref{eq:chisqr}. Now,
the summand no longer depends on $i$ or $k$ explicitly. Thus, the order of the
sums in Eq.~\ref{eq:chisqrequal} may be switched. If we finally interpret $k$
as a spectral energy bin we end up with the goodness as a function of $k$:
\begin{align}
  \begin{split}
  \chi^2_\mathrm{red,comb.}(k) &\propto \sum_{i=1}^N
  \frac{(\mathrm{data}_{i,k} -
    \mathrm{model}_{i,k})^2}{\mathrm{error}_{i,k}^2} \\
  &\propto \sum_{i=1}^N \mathrm{data}_{i,k}^2
  \end{split}
\end{align}
This means that all datasets of the simultaneous fit are first summed up for
each energy bin in the combined reduced $\chi^2$. In contrast to
stacking the data in the first place, however, source variability can still
be taken into account during a simultaneous fit.

Note that once all data-groups have the same number of degrees of
freedom, the $\chi^2_\mathrm{red,comb.}$ (Eq.~\ref{eq:chisqrequal}) is equal
to the classical $\chi^2_\mathrm{red}$ (Eq.~\ref{eq:redchisqr}). To further
investigate the goodness of the simultaneous fit in such a case, the histogram
of the goodness of all data-groups (see Sec.~\ref{sec:histogram}) and, if
possible, the stacked residuals should be investigated.

\begin{figure}
  \centering
  \includegraphics[width=\linewidth]{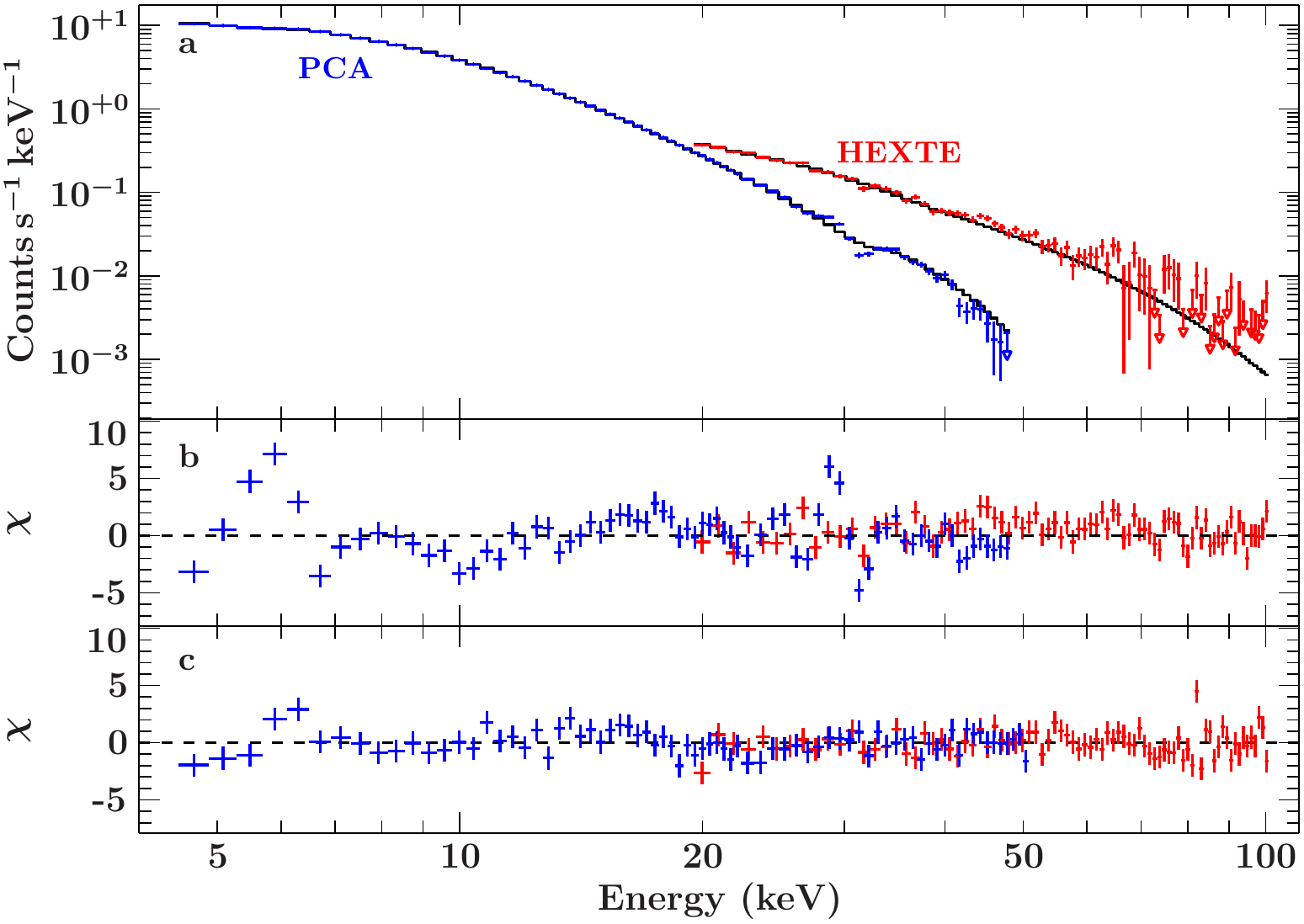}
  \caption{The stacked spectra (a) and stacked residuals of all
  individual data-groups (b) containing 43 \textsl{RXTE}-spectra of
  GRO~J1008$-$57 (blue: PCA; red: HEXTE). Residual features are left in PCA,
  which are caused by calibration uncertainties. These features are not
  detected in that detail in the residuals of the single spectrum with the
  highest signal (c).} \label{fig:1008}
\end{figure}

\section{Examples}
\label{sec:examples}

\subsection{GRO J1008$-$57}
\label{sec:example:gro1008}

The Be X-ray binary GRO J1008$-$57 was regularly monitored by \textsl{RXTE}
during outbursts in 2007 December, 2005 February, and 2011 April with a few
additional pointings by \textsl{Suzaku} and \textsl{Swift}. A detailed
analysis of the spectra has been published in \citet{kuehnel2013a} and in
paper I we have demonstrated, as an example, the advantages of a simultaneous
fit based on these data (see also Fig.~\ref{fig:1008gamma}). The
$\chi^2_\mathrm{red}$ of 1.10 with 3651 degrees of freedom (see Table~4
of K\"uhnel at al., 2013 \citet{kuehnel2013a}) calculated after Eq.~\ref{eq:redchisqr:simfit} indicates a
good fit of the underlying model to the data. Using the combined reduced
$\chi^2$ defined in Eq.~\ref{eq:combredchisqr} we find, however,
$\chi^2_\mathrm{red,comb.} = 1.68$. The reason for this significant worsening
of the goodness are calibration uncertainties in \textsl{RXTE}-PCA, which
are visible in the stacked residuals of all 43 data-groups as shown in
Fig.~\ref{fig:1008}: the strong residuals below 7\,keV are probably caused by
insufficient modeling of the Xe L-edges, the absorption feature at 10\,keV by
the Be/Cu collimator, and the sharp features around 30\,keV by the Xe K-edge
\citep[for a description of the PCA see][]{jahoda2006a}. These calibration
issues have been detected in a combined analysis of the Crab pulsar by
\citet{garcia2014a} as well. However, the calibration issues, which
are responsible for the high $\chi^2_\mathrm{red,comb.}$, do not affect the
continuum model of GRO~J1008$-$57 because of their low significance in the 
individual data-groups. These calibrations features might have an influence,
however, in data with a much higher signal-to-noise ratio than the datasets
used here or once narrow features, such as emission lines, are studied.

\subsection{Vela X-1}

\begin{figure}
  \centering
  \includegraphics[width=\linewidth]{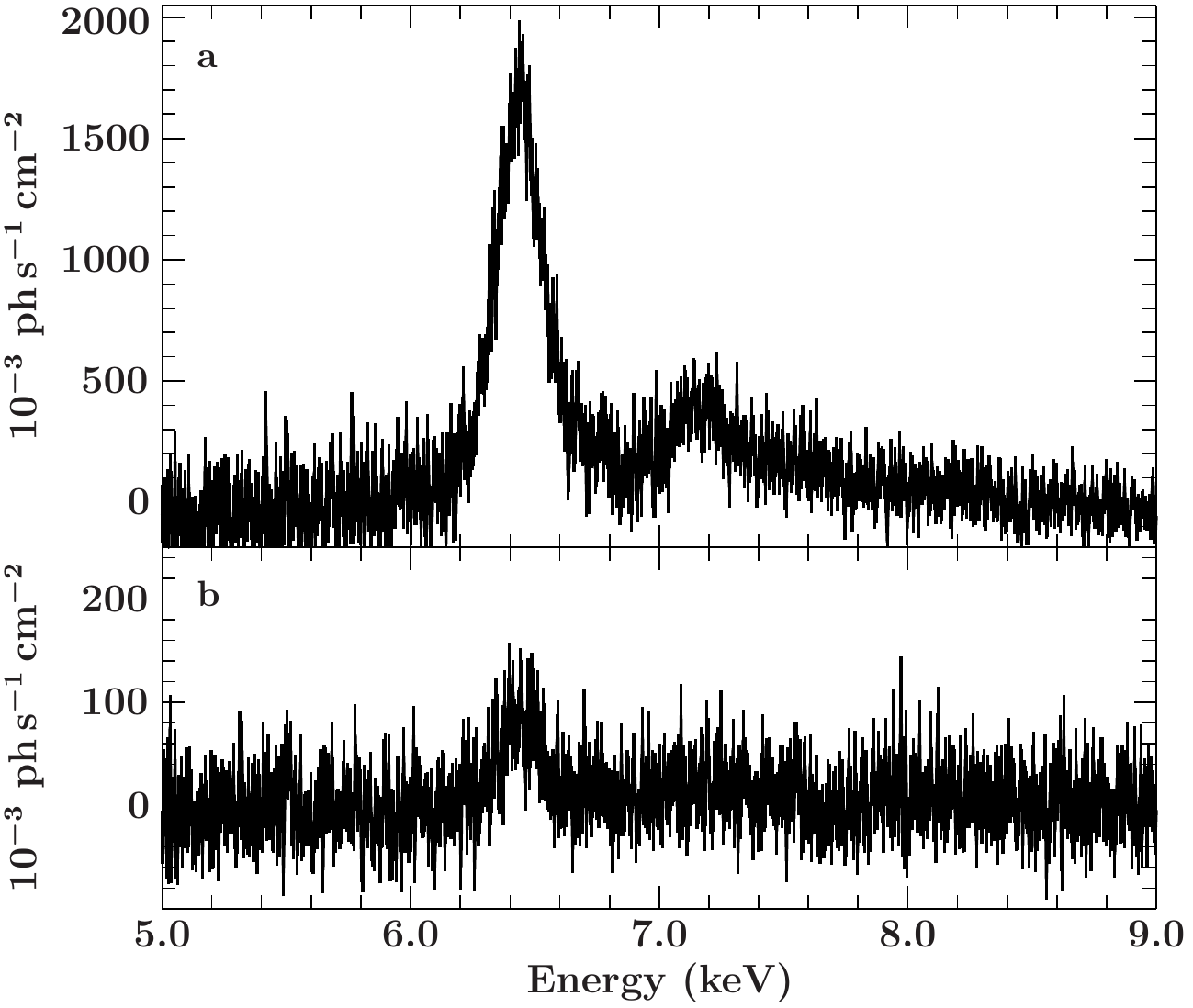}
  \caption{The iron line region of Vela X-1 can be nicely studied in this
  stacked residuals of all 88 \textsl{XMM-Newton}-spectra (a). The model
  includes the continuum shape only and, thus, does not take any fluorescent
  line emission into account. The residuals of the single spectrum with the
  highest signal show the K$\alpha$ emission line only (b). Note that
  the residual flux in this line is $\sim$15 times lower compared to the
  stacked residuals.}
  \label{fig:vela}
\end{figure}

\begin{figure}
  \centering
  \includegraphics[width=\linewidth]{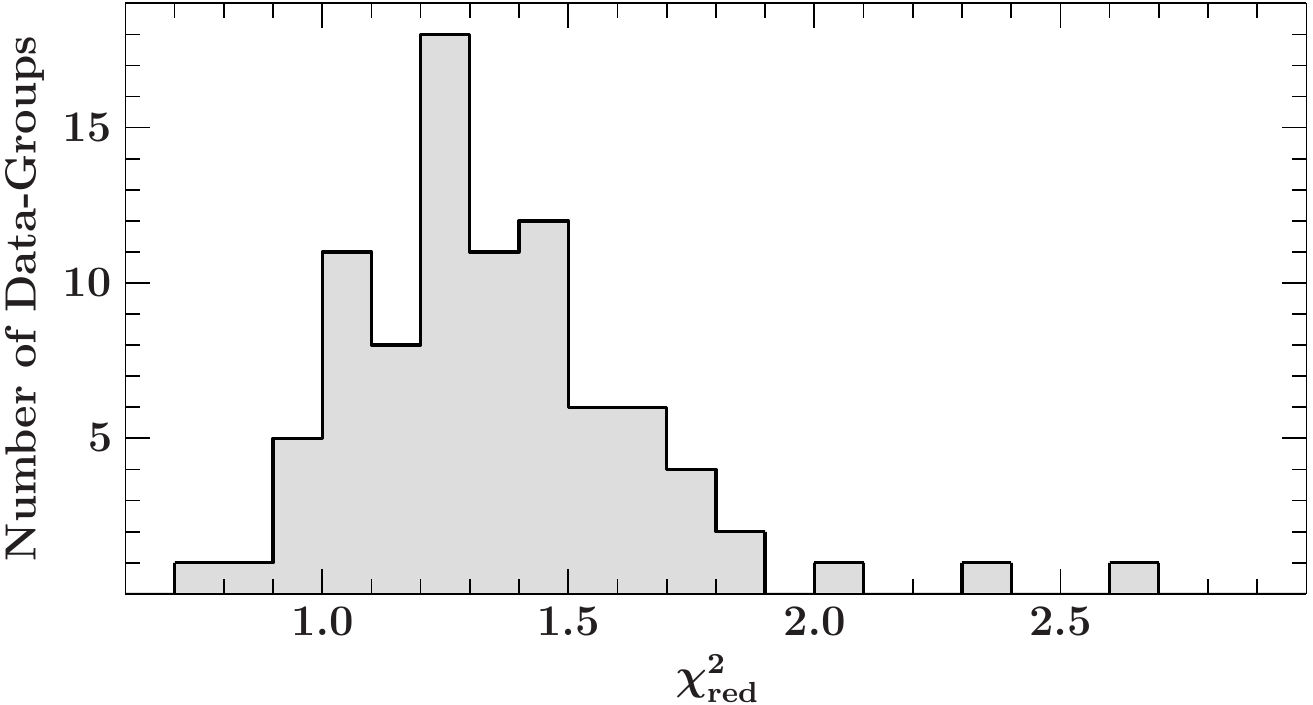}
  \caption{Example for a histogram of the goodness of the fits. Here, the
  distribution of the $\chi^2_\mathrm{red}$ of all individual data-groups
  of the 88 \textsl{XMM-Newton}-spectra of Vela~X-1 are shown.}
  \label{fig:velachi}
\end{figure}

Another excellent example for a simultaneous fit was performed by
\citet{martineznunez2014a}. These authors have analyzed 88 spectra recorded by
\textsl{XMM-Newton} during a giant flare of Vela~X-1. Although the continuum
parameters were changing dramatically within the ${\sim}100$\,ks observation a
single model consisting of three absorbed power-laws is able to describe the
data with a $\chi^2_\mathrm{red} = 1.43$ with 9765 degrees of freedom
\citep{martineznunez2014a}. Due to a global photon index for all power-laws
and data-groups the absorption column densities and iron line fluxes could be
constrained well.

Because every data-group is a single spectrum taken by the \textsl{XMM}-EPIC-PN
camera and a common energy grid was used, the $\chi^2_\mathrm{red,comb.}$
equals the $\chi^2_\mathrm{red}$ here. Thus, it is preferred to calculate the
stacked residuals of all data-groups according to Eq.~\ref{eq:sumresiduals}.
To demonstrate the advantage of this tool we have used the continuum model
only, i.e. without any fluorescence line taken into account, and evaluated
this model without any channel grouping to achieve the highest possible energy
resolution. The resulting stacked residuals in the iron line region
(5--9\,keV) are shown in Fig.~\ref{fig:vela}. The iron K$\alpha$ line at
${\sim}6.4$\,keV and the K$\beta$ line at ${\sim}7.1$\,keV are nicely
resolved. The tail following the K$\beta$ line is probably caused by a slight
mismatch of the continuum model with the data and requires a more detailed
analysis. Note that the flux of this mismatch is a few
$10^{-3}$\,photons\,s$^{-1}$\,cm$^{-2}$, which is detectable only in these
stacked residuals featuring 100\,ks exposure time and after the strong
continuum variability on ks-timescales has been subtracted.

As a further demonstrative example, the histogram of the goodness of the
fits of the 88 data-groups calculated after Eq.~\ref{eq:groupchi} is shown
in Fig.~\ref{fig:velachi}. The median $\chi^2_\mathrm{red}$ is around 1.3,
indicating that the model still could be improved slightly. Investigating
the three outliers with $\chi^2_\mathrm{red} > 2$ indeed proves that the
residuals around in the iron line region are left, which are responsible
for the high $\chi^2_\mathrm{red}$ and very similar to those shown in
Fig.~\ref{fig:vela}. There are no, however, extended residuals visible.
Thus, the continuum parameters presented in \citet{martineznunez2014a} are
still valid.

\subsection{XTE J1859+083}

\begin{figure}
  \centering
  \includegraphics[width=.985\linewidth]{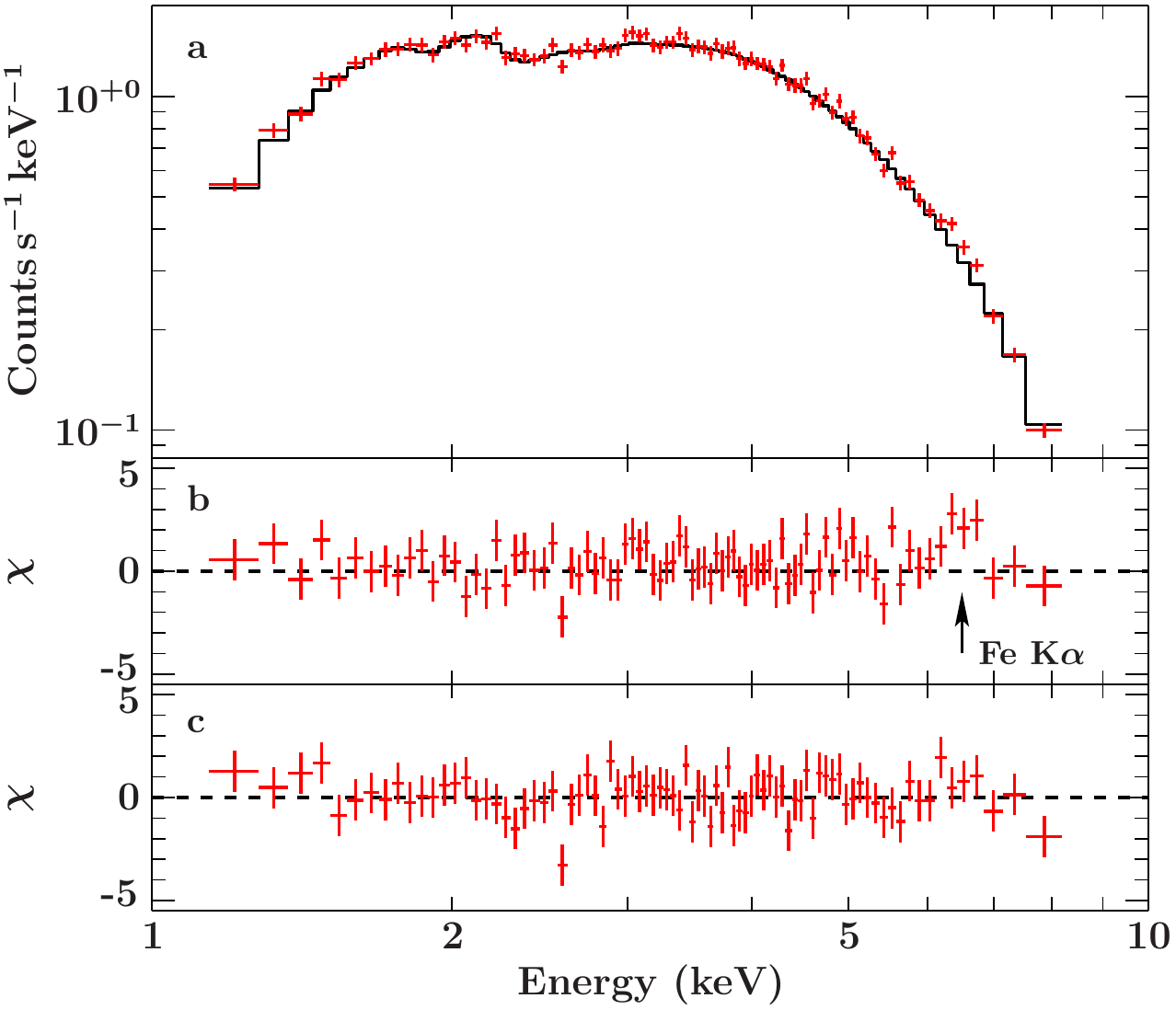}
  \caption{Seven stacked spectra of \textsl{Swift}-observations of
  XTE~J1859+083 (a) and the residuals to the model (b). A weak
  iron K$\alpha$ emission line at 6.4\,keV is visible, which is not detected
  in the residuals of any individual spectrum (c).}
  \label{fig:xte}
\end{figure}

The last example shown in this work is the outburst of the transient pulsar
XTE~1859+083 in 2015 April. This source was in quiescence since its bright
outburst in 1996/1997 \citep{corbet2009a}. During the recent outburst several
short observations by \textsl{Swift} were performed. A first analysis of these
data in combination with \textsl{INTEGRAL} spectra reports an absorbed
power-law shape of the source's X-ray continuum \citep{malyshev2015a}. We have
extracted and analyzed seven \textsl{Swift}-XRT spectra and can confirm these
findings. However, after examining the stacked residuals of all spectra an
iron K$\alpha$ emission line at 6.4\,keV shows up that has not been detected
before (see Fig.~\ref{fig:xte}). We define the equivalent width of this line
as a global parameter and find a value of $60 \pm 40$\,eV (uncertainty is at
the 90\% confidence level, $\chi^2_\mathrm{red,comb.} = 0.98$ with 585 degrees
of freedom).

\section{Summary}

We have continued developing functions to handle simultaneous fits in ISIS,
which we have introduced in paper~I. In particular, we have concentrated on
tools for checking the goodness of the fits to discover failed fits
of individual data-groups or global discrepancies of the model. We
propose to
\begin{itemize}
  \item investigate the distribution of the goodness of fits to all
  individual data-groups
  \item calculate a combined goodness, here the
  $\bf \chi^2_\mathrm{\bf red,comb.}$, which takes the individual nature of
  each data-group into account
  \item look at the stacked residuals of all data-groups to reveal weak
  features
\end{itemize}
during a simultaneous fit in order to find the global best-fit. We
have demonstrated the tremendous benefit of analyzing the stacked
residuals by observations of three accreting neutron stars, in which we
could identify weak features that had not been detected before.

\begin{acknowledgements}
M.~K\"uhnel was supported by the Bundesministerium f\"ur Wirtschaft und
Technologie under Deutsches Zentrum f\"ur Luft- und Raumfahrt grants 50OR1113
and 50OR1207. The \texttt{SLxfig} module, developed by John E. Davis, was used
to produce all figures shown in this paper. We are thankful for the
constructive and critical comments by the reviewers, which were helpful to
significantly improve the quality of the paper.
\end{acknowledgements}

\renewcommand{\bibsep}{0pt}

\end{document}